\documentstyle[preprint,floats,tighten,aps,graphicx]{revtex}

\newif\ifpdf
\ifx\pdfoutput\undefined
\pdffalse % we are not running PDFLaTeX
\else
\pdfoutput=1 % we are running PDFLaTeX
\pdftrue
\fi

\def\gev{{\rm GeV}}
\def\mev{{\rm MeV}}
\def\lqcd{\Lambda_{\rm QCD}}
\def\OMIT#1{{}}

\def\shat{\hat s}
\def\qh{\hat q}
\def\qsh{\hat q^2}
\def\qcut{q_{\rm cut}}
\def\mxcut{m_{\rm cut}}
\def\mxsqcut{\mxcut^2}
\def\xcut{\tilde X}
\def\ycut{\tilde Y}
\def\gcut{G(\qcut^2,\mxcut)}

\def\mbups{m_b^{1S}}

\def\vereq#1#2{\lower3pt\vbox{\baselineskip1pt\lineskip1pt
     \ialign{\\$#1\hfill##\hfil\\$\crcr#2\crcr\sim\crcr}}}

\def\epsblm{\epsilon^2_{\rm BLM}}
\def\d{{\rm d}}

\makeatletter
\def\fmslash{\@ifnextchar[{\fmsl@sh}{\fmsl@sh[0mu]}}
\def\fmsl@sh[#1]#2{%
   \mathchoice
     {\@fmsl@sh\displaystyle{#1}{#2}}%
     {\@fmsl@sh\textstyle{#1}{#2}}%
     {\@fmsl@sh\scriptstyle{#1}{#2}}%
     {\@fmsl@sh\scriptscriptstyle{#1}{#2}}}
\def\@fmsl@sh#1#2#3{\m@th\ooalign{$\hfil#1\mkern#2/\hfil$\crcr$#1#3$}}

\makeatother
\begin{document}
%%%%%%%%%%%%%%%%%%%%%%%%%%%%%%%%%%%%%%%%%%
%Some more stuff to get graphics to work
\ifpdf
\DeclareGraphicsExtensions{.pdf, .jpg}
\else
\DeclareGraphicsExtensions{.eps, .jpg}
\fi

%%%%%%%%%%%%%%%%%%%%%%%%%%%%%%%%%%%%%%%%%%
%Define Title, Author, Address, Preprint#

\preprint{\vbox{ \hbox{UTPT-01-08}\hbox{UCSD/PTH 01-05}
   \hbox{LBNL-47800} \hbox{hep-ph/0107074}  \hbox{} }}

\title{Precision determination of $|V_{ub}|$ from inclusive decays}
\author{Christian W.\ Bauer,$^{a}$
   Zoltan Ligeti,$^{b}$ and Michael Luke$^{b,c}$}

\address{ \vbox{\vskip 0.truecm}
   $^a$Physics Department, University of California at San
Diego, La Jolla, CA 92093 \\[8pt]
$^b$Ernest Orlando Lawrence Berkeley National Laboratory\\
University of California, Berkeley, CA 94720 \\[8pt]
$^c$Department of Physics, University of Toronto, \\
     60 St.\ George Street, Toronto, Ontario, Canada M5S 1A7 \\ [8pt]}

\maketitle

\begin{abstract}%
We propose determining $|V_{ub}|$ from inclusive semileptonic $B$ decay using
combined cuts on the leptonic and hadronic invariant masses to eliminate the
$b\to c$ background.  Compared to a pure dilepton invariant mass cut, the
uncertainty from unknown order $\lqcd^3/m_b^3$ terms in the OPE is
significantly reduced and the fraction of $b\to u$ events is roughly doubled.
Compared to a pure hadronic invariant mass cut, the uncertainty from the
unknown light-cone distribution function of the $b$ quark is  significantly
reduced.  We find that $|V_{ub}|$ can be determined with theoretical
uncertainty at the 5--10\% level.

\end{abstract}

\newpage

\section{Introduction}

The magnitude of the Cabibbo-Kobayashi-Maskawa matrix element $V_{ub}$  is an
important ingredient in overconstraining the unitarity triangle by  measuring
its sides and angles.   Inclusive semileptonic $b\to u$ decay provides  the
theoretically cleanest method of measuring $|V_{ub}|$ at present, since  it can
be calculated model independently using an operator product expansion  (OPE) as
a double expansion in powers of $\lqcd/m_b$ and $\alpha_s(m_b)$ \cite{OPE}.
However, the phase space cuts which are required to eliminate the overwhelming
background from $b\to c$ decay typically cause the standard OPE to fail.  This
is the case both for the cut on the charged lepton energy, $E_\ell >
(m_B^2-m_D^2) / 2m_B$ \cite{Elexp}, as well as for the cut on the  hadronic
invariant mass, $m_X < m_D$ \cite{mXcut,mXold,mXexp}. In both of these cases,
the  standard OPE becomes, in the restricted region, an expansion in powers of
$\lqcd m_b/m_c^2$, which is of order unity.

Recently we showed that a cut on the dilepton invariant mass can be used  to
reject the background from $b \to c$ decay \cite{BLL,BLL2}, while still
allowing an expansion in local operators.  Imposing a cut $q^2 > (m_B-
m_{D})^2$ (where $q$ is the four-momentum of the virtual $W$) removes the $b
\to c$ background while leaving the OPE valid.  This approach has the advantage
of being model independent, but is only sensitive to $\sim 20\%$ of the rate,
as opposed to $\sim 80\%$ for a $m_X < m_D$ hadronic invariant mass cut.
Besides the sensitivity to $m_b$, the main uncertainty in the analysis using a
pure $m_X$ cut comes from uncalculable corrections, formally of order
$\lqcd/m_b$, to the $b$ quark light-cone distribution function,\footnote{This
assumes that the light-cone distribution function of the $b$ quark is
determined from the $B\to X_s\gamma$ photon spectrum; otherwise the model
dependence is formally $O(1)$.} while in the case of the pure $q^2$ cut from
the order $(\lqcd/m_b)^3$ corrections in the OPE, the importance of which was
recently stressed \cite{Voloshin}.  In addition, because of finite detector
resolution, the actual experimental cut on $q^2$ may be larger than the optimal
value of $(m_B-m_D)^2$, and the theoretical error in $|V_{ub}|$ grows rapidly
as $q^2$ is raised.

In this paper we propose to improve on both methods by combining cuts on the
leptonic and hadronic invariant mass.  Varying the $q^2$ cut in the  presence
of a cut on $m_X$ allows one to interpolate continuously between the limits  of
a pure $q^2$ cut and a pure $m_X$ cut.  We examine how a combined cut on $m_X$
and $q^2$ can minimize the overall uncertainty.  This also allows a precision
determination of $|V_{ub}|$ to be obtained with cuts which are away from the
threshold for $B\to X_c\ell\bar\nu_\ell$, an important criterion for realistic
detector resolution.

In Sec.~\ref{kinsection} we discuss the regions of phase space and  explain
which ones are accessible within the standard OPE.   In  Sec.~\ref{combinedcut}
we present the decay rate with a combined cut on the leptonic and  hadronic
invariant mass to order $\lqcd^2/m_b^2$ in the OPE and to order  $\alpha_s^2
\beta_0$ in the perturbative expansion, including a detailed investigation of
the theoretical uncertainties.  Our results  are summarized in
Sec.~\ref{results}.

\section{Kinematics}\label{kinsection}

\begin{figure}[t]
\centerline{\includegraphics[width=4.5in]{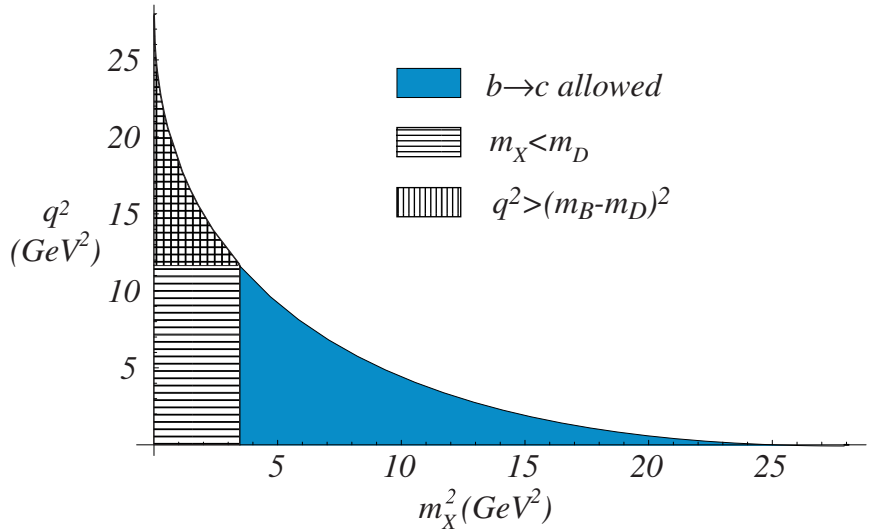}}
\caption{The Dalitz plot for $b\to u$ semileptonic decay, indicating the
regions corresponding to $b\to c$ decay (shaded), the lepton invariant  mass
cut $q^2 > (m_B-m_D)^2$ (vertically striped), and the hadron invariant mass
cut $m_X< m_D$ (horizontally striped).}
\label{dalitz1}
\end{figure}

The Dalitz plot for $b\to u$ semileptonic decay in the $q^2-m_X^2$ plane  is
shown in Fig.~\ref{dalitz1}.  While the region of phase space contained  by the
$q^2 > (m_B-m_D)^2$ cut corresponds to a subset of the region $m_X<m_D$,  the
theoretical prediction for the former region is better behaved  \cite{BLL}.
This may seem counterintuitive, since uncertainties for inclusive  observables
usually decrease the more inclusive the quantity is.  The present  situation
occurs because the OPE breaks down when the kinematics is restricted to  large
energy and low invariant mass final states, for which $m_X^2/E_X \sim  \lqcd$.
As it is explained below, this kinematics dominates the lower left  corner of
the Dalitz plot in  Fig.~\ref{dalitz1}, and that is why the OPE is better
behaved in the restricted region determined by the $q^2$ cut.

More precisely, there are three distinct regions of phase space, in which the
behavior of the OPE is qualitatively different.  Over most of the Dalitz plot,
the kinematics typically satisfies
\begin{equation}
   m_X \gg \lqcd\,,  \qquad  m_X^2/E_X \gg \lqcd \,,
\end{equation}
and the inclusive rate may be expanded in powers of $\lqcd/m_b$ via the OPE.
The leading order term is the $b$ quark decay result, and the higher order
terms are parametrized by matrix elements of local operators.  This is the
simplest region theoretically, since reliable predictions can be made knowing
only the first few matrix elements, which may be determined from other
processes.  The situation is more complicated in the ``shape function" region,
which is dominated by low invariant mass and high energy final states
\begin{equation}\label{shaperegion}
   m_X \gg \lqcd\,,  \qquad  m_X^2/E_X \sim O(\lqcd) \,.
\end{equation}
In this region, a class of contributions proportional to powers of $\lqcd
E_X/m_X^2$ must be resummed to all orders.  The OPE is replaced by a  twist
expansion, in which the leading term depends on the light-cone  distribution
function of the $b$ quark in the $B$ meson. Since this is a  nonperturbative
function, the leading order prediction is model dependent, unless the
distribution function is measured from another process.  Even if this  light
cone distribution function is extracted from the photon energy spectrum  in $B
\to X_s \gamma$ \cite{structure,extractshape}, the unknown higher order
corrections are only suppressed by $\lqcd/m_b$.  Finally, in the  resonance
regime,
\begin{equation}\label{resonance}
   m_X \sim O(\lqcd) \,,
\end{equation}
the final state is dominated by a few exclusive resonances and the  inclusive
description breaks down.  In this case neither the local OPE nor the twist
expansion is applicable.

Which of these situations applies to the kinematic regions $m_X < m_D$ 
and $q^2 > (m_B-m_D)^2$ depends on the relative sizes of $m_b$, $m_c$ and  $\lqcd$. It
seems most reasonable to treat
\begin{equation}
   \lqcd\, m_B \sim O(m_D^2)\,,
\end{equation}
since neither side is much larger than the other. Cutting only on the hadronic
invariant mass (or on $E_\ell$), the  hadronic energy can extend all the way to
order $m_B$,
\begin{equation}\label{invtmassphase}
   m_X \sim m_D\,,  \qquad   E_X \sim m_B\,,
\end{equation}
and so $m_X^2/E_X$ is typically of order $\lqcd$.
By contrast, the cut on $q^2$ implies
\begin{eqnarray}\label{q2phase}
m_X \sim m_D\,, \qquad
   E_X = \frac{m_B^2 -q^2 + m_X^2}{2m_B} \sim m_D\,,
\end{eqnarray}
and so typically $m_X^2/E_X \sim m_D$. Viewing $m_D\gg\lqcd$, both  regions are
parametrically far from the resonance regime (\ref{resonance}).   However, the
$m_X < m_D$ (or $E_\ell > (m_B^2-m_D^2) / 2m_B$) region is in the shape
function regime [see, Eq.~(\ref{shaperegion})], and thus sensitive to  the
light-cone distribution function.  In contrast, the region $q^2 > (m_B-m_D)^2$
is parametrically far from both the resonance and shape function regimes.

Thus, the cut on $q^2$ eliminates the region where the structure function is
important, making the calculation of the partially integrated rate possible in
an expansion of local operators.  However, from Eq.\ (\ref{q2phase}), imposing a
cut $q^2<\qcut^2=(m_B-m_X)^2$ results in the effective expansion parameter for
the OPE being
\begin{equation}\label{effexp}
{\lqcd E_X\over m_X^2}\sim {\lqcd\over m_X}\sim {m_b\lqcd\over m_b^2-\qcut^2},
\end{equation}
and so the convergence of the OPE gets worse as $\qcut^2$ is raised.
For $\qcut^2 = (m_B-m_D)^2 \simeq
(m_b-m_c)^2$, the OPE is an expansion in $\lqcd/m_c$ \cite{neubertq2}.
For a very high cut on $q^2$ (say,  above
$\sim 18\,\mbox{GeV}^2$), the phase space is restricted to the resonance
region, causing a breakdown of the OPE.

For the pure $q^2$ cut, the largest uncertainties originate from the $b$  quark
mass and the unknown contributions of dimension-six operators,  suppressed by
$[m_b\lqcd/(m_b^2-\qcut^2)]^3$.  In this paper we propose that the
uncertainties can be reduced considerably by lowering the cut on $q^2$  below
$(m_B-m_D)^2$, and using a simultaneous cut on $m_X$ to reject $b \to c$
events.  It is obvious that lowering $\qcut^2$ all the way to zero would
result in the rate with just the cut on $m_X$, which depends strongly on the
light-cone distribution function.  Thus lowering $\qcut^2$ in the presence of
a fixed cut on $m_X$ increases the uncertainty from the structure function,
while decreasing the uncertainty from the matrix elements of the dimension-six
operators.  The optimal combination of the two cuts is somewhere in  between
the pure $q^2$ and pure $m_X$ cuts.  In the rest of this paper we calculate
the the partially integrated rate and its uncertainty in the presence of cuts
on  $q^2$ and $m_X$.

\section{Combined Cuts}\label{combinedcut}

The integrated rate with a lower cut $\qcut^2$ on $q^2$ and an upper cut
$\mxcut$ on $m_X$ may be written as
\begin{equation}\label{defineg}
\int_{\hat\qcut^2}^1 {\rm d}\hat q^2 \int_0^{\hat s_0}
   {\rm d}\hat s\, {{\rm d}\Gamma\over {\rm d}\hat q^2 {\rm d}\hat s}
\equiv {G_F^2 |V_{ub}|^2\, (4.7\,\gev)^5\over 192\pi^3}\; \gcut\,,
\nonumber
\end{equation}
where where $\hat q = q/m_b$, $\shat = (v - \hat q)^2$ is the rescaled
partonic invariant mass, $v$ is the four-velocity of the decaying $B$ meson,
and
\begin{eqnarray}\label{s0limit}
\hat s_0 = \cases{
\displaystyle \left( 1 - \sqrt{\qsh}\right)^2
   & for~ $\mxcut > m_B - m_b\,\sqrt{\qsh}$\,, \vspace*{6pt}\cr
\displaystyle 0
   & for~ $\mxcut^2 < (m_B-m_b\,\qsh)\, (m_B-m_b)$\,, \vspace*{6pt}\cr
\displaystyle \frac\mxsqcut{m_B m_b} + \left( \frac{m_B}{m_b} - 1 \right)
   \left( \frac{m_b}{m_B}\,\qsh - 1 \right)
   & otherwise\,. \cr}
\end{eqnarray}
The hadronic invariant mass $m_X$ is related to $\hat q^2$ and $\hat s$ by
\begin{equation}\label{rel}
  m_X^2 = \shat\, m_B m_b  + (m_B-m_b)(m_B- \qsh m_b) \,.
\end{equation}
$\gcut$ is the ratio of the semileptonic $b\to u$ width with cuts on $q^2$ and
$m_X$ to the full width at tree level with $m_b=4.7\,\gev$.  The fraction of
semileptonic $b\to u$ events included in the cut rate is $\simeq 1.21\, \gcut$.
Note that  the
$m_b^5$ prefactor, a large source of uncertainty, is included in $\gcut$. The
theoretical uncertainty in the extraction of $|V_{ub}|$ is therefore  half the
uncertainty in the prediction for $\gcut$.

\subsection{Standard OPE}

For $q^2 > (m_B-\mxcut)^2$, the effects of the structure function are
parametrically suppressed, and correspond to including a class of  subleading
higher order terms in the OPE.  In this region the standard OPE is
appropriate, and the double differential decay rate is given by
\begin{eqnarray}\label{q2spec}
{1\over \Gamma_0}\, {{\rm d}\Gamma \over {\rm d}\qsh {\rm d}\shat} &=&
   \delta(\shat) \left[ \bigg( 1 + {\lambda_1\over 2m_b^2} \bigg)\, 2\,
   (1-\qsh)^2\, (1+2\qsh) + {\lambda_2\over m_b^2}\, (3 - 45\hat q^4 +
   30\hat q^6) \right] \nonumber\\
&&{} + {\alpha_s(m_b) \over \pi}\, X(\qsh,\shat)
   + \bigg( {\alpha_s(m_b) \over \pi} \bigg)^2\, \beta_0\, Y(\qsh,\shat)
   + \ldots \,,
\end{eqnarray}
where $\beta_0 = 11 - 2n_f/3$ and
\begin{equation}
  \Gamma_0 = {G_F^2\, |V_{ub}|^2\, m_b^5 \over 192\, \pi^3}
\end{equation}
is the tree level $b\to u\ell\bar\nu$ decay rate.  The matrix element
$\lambda_2$
is known from the $B^*-B$ mass splitting, $\lambda_2=0.12\,\gev^2$ (the
uncertainty
in this relation is included in the $O(1/m_b^3)$ terms).  $\lambda_1$ is much
less
well known but, as is clear from (\ref{q2spec}), the rate is very insensitive to
it.  The ellipses in
Eq.~(\ref{q2spec}) denote order $\alpha_s^2$ terms not enhanced by  $\beta_0$,
order $(\lqcd/m_b)^2$ terms proportional to derivatives of  $\delta(\shat)$,
and higher order terms in both series. The function $X(\qsh,\shat)$ can be
obtained from the triple differential rate given in \cite{defazio}, and the
function $Y(\qsh,\shat)$ was calculated numerically in \cite{LSW}.

\begin{figure}[t]
\centerline{\includegraphics[width=4in]{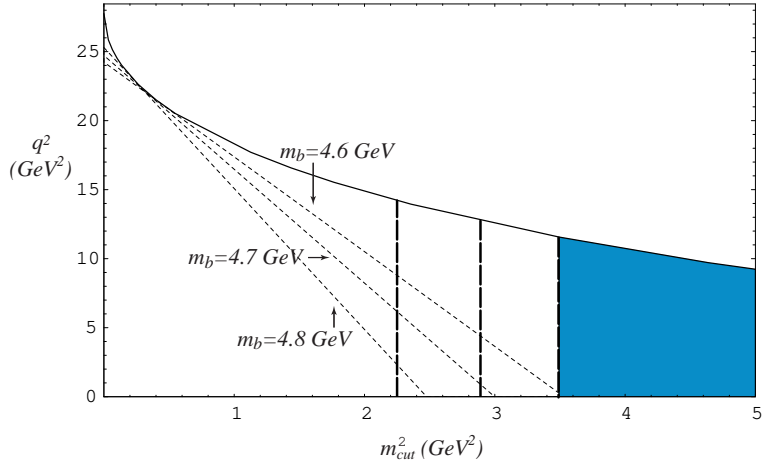}}
\caption{The thin dashed lines show the location of the perturbative
singularity of ${\rm d}\Gamma_c(\mxcut)/{\rm d}q^2$, given by
Eq.~(\ref{qsqblowup}), for $m_b =4.6,\ 4.7$ and $4.8\,\gev$.  The thick dashed
lines correspond to $\mxcut=1.5,\ 1.7$ and $1.86\,\gev$.  The intersection of
the thick and thin dashed lines give qualitatively, for  a given value of
$\mxcut$, the value of $\qcut^2$ below which the effects of the distribution
function become large.}
\label{singularityplot}
\end{figure}

The perturbative contributions to the differential rate in  Eq.~(\ref{q2spec})
are finite for $\shat>0$, where only bremsstrahlung diagrams contribute,  but
singular as $\shat\to 0$.  For a fixed value of $m_X$, setting $\shat =  0$ in
Eq.~(\ref{rel}) determines how far $q^2$ can be lowered without  encountering
the singularity.  Since the singularity is smoothed out by the $b$ quark
light-cone distribution function, such low values of $q^2$ correspond to  the
shape function region.  Throughout this paper we will therefore stay  away from
this region by only considering values of $\qcut^2$ and $\mxcut$  satisfying
\begin{equation}\label{qsqblowup}
   \qcut^2 > m_B\, m_b - \mxsqcut\,\frac{m_b}{m_B-m_b} \,.
\end{equation}
This is illustrated in Fig.~\ref{singularityplot}.  Note that if  $\mxcut$ is
lowered, $\qcut^2$ must be increased to keep the uncertainty at a roughly
constant level.  If the difference between the left- and right-hand  sides of
Eq.~(\ref{qsqblowup}) is at least few times $\lqcd\,m_b$ then we are far  from
the shape function region, and the OPE is well behaved.  In this case  the tree
level result is not sensitive to the cut on $m_X$, and the $\qsh$  spectrum
including a hadronic invariant mass cut is given by
\begin{eqnarray}\label{singlecut}
{1\over \Gamma_0}\, {{\rm d}\Gamma_c(\mxcut)\over {\rm d}\qsh}
&=& \bigg( 1 + {\lambda_1\over 2m_b^2} \bigg)\, 2\, (1-\qsh)^2\,
(1+2\qsh)
   + {\lambda_2\over m_b^2}\, (3 - 45\hat q^4 + 30\hat q^6) \nonumber\\
&&{} + {\alpha_s(m_b) \over \pi}\, \xcut(\qsh,\mxcut)
   + \bigg( {\alpha_s(m_b) \over \pi} \bigg)^2\, \beta_0\,
\ycut(\qsh,\mxcut)
   + \ldots \,,
\end{eqnarray}
where the functions $\xcut(\qsh,\mxcut)$ and $\ycut(\qsh,\mxcut)$ are  given in
the Appendix.

The differential decay rate in Eq.~(\ref{singlecut}) is given in terms of the
pole mass, $m_b^{\rm pole}$.  It is well-known that use of the  pole mass
introduces spurious poor behavior of the perturbation series.  Although  this
cancels in relations between physical observables, it is simplest to  avoid it
from the start by using a better  mass definition. There are a number of
possibilities; here we choose the $1S$ mass, which is defined as one  half of
the $\Upsilon(1S)$ mass in perturbation theory.  To the
order we are working, it is related to the pole mass by
\begin{eqnarray}\label{mb1s}
\mbups = m_b^{\rm pole} \left\{1-\frac{\left(C_F\, \alpha_s\right)^2}{8}
   \left[1\epsilon + \frac{\alpha_s}{\pi}\, \beta_0
   \left(\ln{\mu\over m_b \alpha_s C_F}+\frac{11}{6}\right) \epsilon_{\rm BLM}^2
+ \ldots \right]
\right\} ,
\end{eqnarray}
where powers of $\epsilon \equiv 1$ count the order in the upsilon
expansion\cite{upsexp},  $C_F=4/3$,
and $\epsblm$ denotes the ``BLM-enhanced" (by a factor of $\beta_0$)
$O(\epsilon^2)$ term.
Terms of order $\alpha_s^n$ in Eq.~(\ref{singlecut}) should be counted as order
$\epsilon^n$, and terms of the same order in $\epsilon$ in the two series
should be combined.  The mismatch in orders of $\alpha_s$ between
(\ref{singlecut}) and  (\ref{mb1s}) is required for the bad behavior of  the
two series to cancel\cite{upsexp}.

The uncertainties in the OPE prediction for $\gcut$ from Eq.\ (\ref{singlecut})
come
from three separate sources: perturbative uncertainties from the unknown full
two-loop result, uncertainties in $b$ quark mass and uncertainties due to
unknown matrix  elements of local operators at $O(1/m_b^3)$ in the OPE.   In the
following
subsections we will estimate each of these uncertainties separately as the
fractional  errors on
$\gcut$.  The fractional uncertainty in $|V_{ub}|$ then is one half of the
resulting value.

\subsubsection{Perturbative uncertainties}\label{uncert_pert}

\begin{figure}[t]
\centerline{\includegraphics[width=6in]{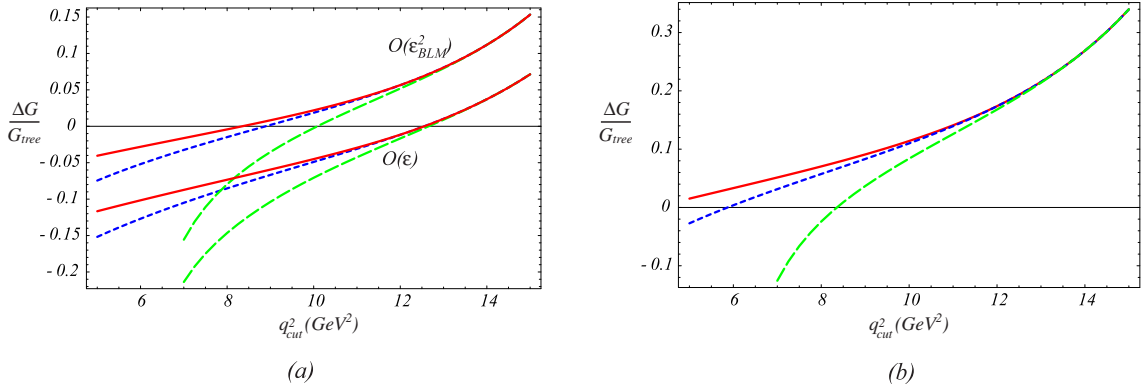}}
\caption{(a) The $O(\epsilon)$ and  $O(\epsblm)$ contributions to
$\gcut$
(normalized to the tree level result) for hadronic invariant mass cut
$\mxcut=1.86\,\gev$ (solid lines), $1.7\,\gev$ (short dashed lines) and
$1.5\,\gev$ (long dashed lines).  (b) Scale variation of the perturbative
corrections: The difference between the perturbative
corrections to $\gcut$, normalized to the tree level
result, for $\mu=4.7\,\gev$ and $\mu=1.6\,\gev$.}
\label{perturbative}
\end{figure}

The relative sizes of the $O(\epsilon)$ and $O(\epsblm)$ corrections to $\gcut$
are plotted in Fig.~\ref{perturbative}(a), for $\mu=4.7\,\gev$.  We note that
for a given value of $\mxcut$, the perturbation series is poorly behaved for
$\qcut^2$ both larger and smaller than some optimal range. For large $\qcut^2$,
this behaviour arises because the invariant mass of the final hadronic state is
constrained to be small, and so perturbation theory breaks down.  For lower
values of $\qcut^2$, the perturbative singularity discussed in the previous
section is being approached, and there are large Sudakov logarithms which blow
up. These Sudakov logarithms may in principle be resummed, but since our point
in this paper is to avoid the shape function region entirely, we will stay in
the intermediate region where ordinary perturbation theory is well behaved.

We may estimate the error in the perturbation series in two ways: (a) as the
same size as the last term computed, the order $\epsblm$ term, or (b) as the
change in the perturbation series by varying $\mu$ over some reasonable range.
These are illustrated in Fig.~\ref{perturbative} (a) and (b), respectively. In
Fig.~\ref{perturbative}(b) we vary the renormalization scale between
$\mu=4.7\,\gev$ and $\mu=m_b/3\sim 1.6\,\gev$, and plot the change in the
perturbative result (including both $O(\epsilon)$ and $O(\epsblm)$ terms).  For
a given set of $\qcut^2$ and $\mxcut$, we take the perturbative error to be the
larger of (a) and (b).

Note that since both the $O(\epsilon)$ and $O(\epsblm)$ terms change sign in the
region of interest, this approach may underestimate the error in the
perturbative series, particularly near the values of the cuts where the
$O(\epsilon^2 \beta_0)$ term or the scale variation vanishes.  To put the estimate
of the perturbative uncertainty on firmer grounds, a complete two-loop
calculation of the double  differential rate, ${\rm d}\Gamma/{\rm d}q^2{\rm
d}m_X$, is most desirable. This is one of the ``simpler" two-loop calculations,
since the phase space of the  leptons can be factorized.

As an alternate approach, Refs.\ \cite{neubertq2,neubbech} use the
renormalization group to sum leading and subleading logarithms of
$m_b/(m_b-\sqrt{\qcut^2})$ (for a pure $q^2$ cut). However, since this log is
not large in the regions we are considering, it is not clear that this improves
the result.  For example, resumming  leading logs of $m_c/m_b$ for $B\to D^*$
semileptonic decay at zero recoil in HQET is known to provide a poor
approximation to the full two-loop result, and including the power suppressed
$(m_c/m_b)\, \alpha_s^n\ln^n (m_c/m_b)$ terms makes the agreement even worse
\cite{hqetlogs}.

\subsubsection{Uncertainties in the $b$ quark mass}\label{uncert_mass}

\begin{figure}[t]
\centerline{\includegraphics[width=6in]{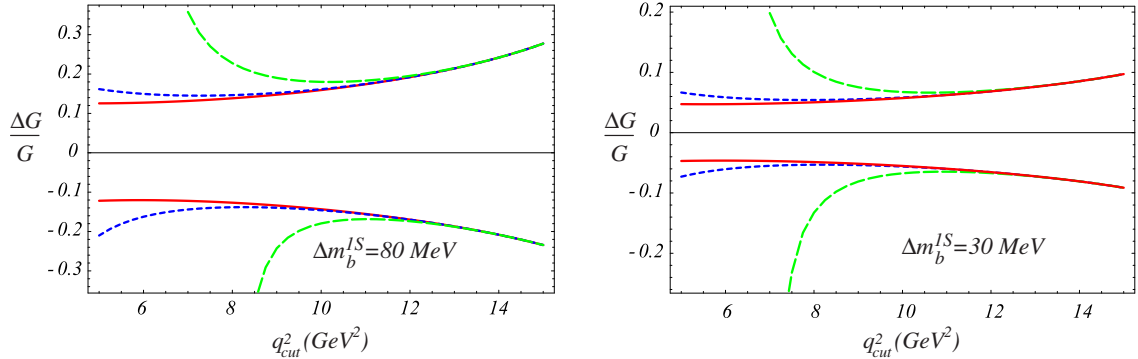}}
\caption{The fractional effect of a $\pm 80\,\mev$ and $\pm 30\,\mev$
uncertainty in $\mbups$ on $\gcut$ for $\mxcut=1.86\,\gev$ (solid line),
$1.7\,\gev$ (short dashed line) and $1.5\,\gev$ (long dashed line).}
\label{mass}
\end{figure}

The partially integrated rate depends sensitively on the value of the $b$ quark
mass due both to the $m_b^5$ factor in $\gcut$ and the cut on $q^2$, as stressed
in \cite{neubertq2}.
Currently, the smallest error of the $1S$ mass is quoted from sum rules
\cite{sumrules1,benekesigner,hoang00}. Ref.~\cite{hoang00} obtains the value
$\mbups = 4.69\pm0.03\,\gev$ by fitting an optimized linear combination of
moments of the $e^+e^-\to b\,\bar b$ spectrum, which may underestimate the
theoretical error \cite{benekesigner}; the authors of \cite{benekesigner} cite a
 similar central value with a more conservative error of $\pm  0.08\,\gev$. In
Fig.~\ref{mass} we show the effects of a $\pm 80\,\mev$ and a $\pm 30\,\mev$
uncertainty in $\mbups$ on $\gcut$, using the central value $\mbups =
4.7\,\gev$.  The  latter error may be achievable using moments of various
$B$ decay distributions \cite{decaydists}.

\subsubsection{${\cal O}(\lqcd^3/m_b^3)$ uncertainties}\label{uncert_m3}

As discussed in Section II, the convergence of the OPE gets worse as $\qcut^2$
is raised.  Since the contribution from $\lambda_1$ in the OPE is small for all
values of $\qcut^2$ (see (\ref{singlecut})) and $\lambda_2$ is known, the
largest uncertainty from unknown nonperturbative terms in the OPE arises at
$O(\lqcd^3/m_b^3)$ \cite{m3}. The effects of these terms were estimated in
\cite{BLL} by varying the values of the corresponding matrix elements over the
range expected by dimensional analysis, and determining the corresponding
uncertainty in $|V_{ub}|$ as a function of $\qcut^2$.  Since the $b$ quark decay
result at tree level is insensitive to the cut on $m_X$, as long as $\mxcut$ is
not too low, these results may be immediately taken over to the present
analysis. However, the cut on $m_X$  allows $\qcut^2$ to be lowered below
$(m_B-m_D)^2$, resulting in a significant reduction of the uncertainty, since by
(\ref{effexp}) it scales as $[m_b\lqcd / (m_b^2-\qcut^2)]^3$.

In addition to these corrections, Voloshin \cite{Voloshin} has recently stressed
the importance of the contribution from weak annihilation (WA) (this uncertainty
was included but underestimated in \cite{BLL}).  WA arises at $O(\lqcd^3/m_b^3)$
in the OPE, but is enhanced by a factor of $\sim16\pi^2$ because there are only
two particles in the final state compared with $b\to u\ell\bar\nu_\ell$. 
Because WA contributes only at the endpoint of the $q^2$ spectrum, it is
independent of $\qcut^2$ and $\mxcut$:
\begin{equation}\label{WAcont}
{\d\Gamma_{WA}\over \d q^2}=-{2 G_F^2 |V_{ub}|^2 m_b^2\over 3\pi}\delta(q^2-m_b^2)
{1\over 2 m_B}\langle B|O^u_{V-A}-O^u_{S-P}|B\rangle
\end{equation}
where
\begin{equation}
O^q_{V-A}=\textstyle{1\over 4}\bar h_b\gamma_\mu(1-\gamma_5) q
\bar q\gamma^\mu(1-\gamma_5) h_b,\quad
O^q_{S-P}=\textstyle{1\over 4}\bar h_b(1-\gamma_5) q
\bar q(1-\gamma_5) h_b.
\end{equation}
The matrix element in (\ref{WAcont}) vanishes for both charged and neutral $B$'s
under the factorization hypothesis (in which case it corresponds to pure
annihilation, which vanishes by helicity for massless leptons), and so the size
of the WA effect depends on the size of factorization violation.  Following the
discussion in \cite{Voloshin} we define the bag constants $B_i$ by
\begin{equation}
{1\over 2 m_B}\langle B|O^u_{V-A}|B \rangle\equiv {f_B^2 m_B\over 8} B_1,\quad
{1\over 2 m_B}\langle B|O^u_{S-P}|B \rangle\equiv {f_B^2 m_B\over 8} B_2.
\end{equation}
Under factorization, $B_1=B_2=1$ for $B^\pm$, and $B_1=B_2=0$ for $B_d$, while
Ref.~\cite{Voloshin} suggests a 10\% violation of factorization, $|B_1-B_2|\sim
0.1$, as being a reasonable estimate. This gives a constant shift to $\gcut$ of
\begin{equation}
\delta\gcut=16\pi^2(B_2-B_1) {f_B^2\over m_b^2}\sim 0.03\left({f_B\over
0.2\,\gev}
\right)^2\left({B_2-B_1\over 0.1}\right).
\end{equation}
While this corresponds to only a $\sim 3\%$ correction to the total $b\to
u\ell\bar\nu_\ell$ rate, the importance of this correction grows as the cuts
reduce the number of events.\footnote{Note that by the same token, this implies
a $\sim15\%$ uncertainty in $|V_{ub}|$ extracted from the charged lepton energy
endpoint region \cite{structure,leptonendpoint1,leptonendpoint2}, $E_\ell >
(m_B^2-m_D^2) / 2m_B$, even when the light-cone distribution function of the $b$
quark is determined from $B\to X_s\gamma$.}

\begin{figure}[t]
\centerline{\includegraphics[width=4in]{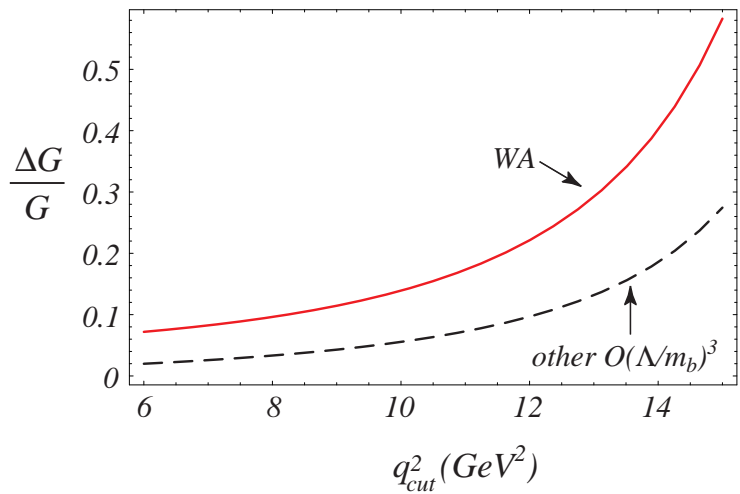}}
\caption{Estimate of the uncertainties due to dimension-six terms in the OPE as
a function of $\qcut^2$ from weak annihilation (WA) (solid line) and other
operators (dashed line).}
\label{mcubed}
\end{figure}

The estimated uncertainty from these two classes of $\lqcd^3/m_b^3$ corrections
to $\gcut$ are plotted in Fig.~\ref{mcubed}, for $B_2-B_1=0.1$.  Since the
uncertainty from WA is roughly a factor of two larger than from the other terms,
we use the estimate from WA to determine the theoretical error on $\gcut$ from
$1/m_b^3$ effects.

The effects of WA are particularly difficult to estimate because they arise from
a small matrix element (factorization violation) multiplying a large coefficient
($16\pi^2$), and so further experimental input is required to have confidence in
this error estimate.  Such spectator effects could be computed using lattice
QCD, or could be constrained experimentally from the difference of $|V_{ub}|$
extracted from neutral and charged $B$ decay, or from an experimental
measurement of the difference of the semileptonic widths of the $D^0$ and $D_s$
\cite{Voloshin}.

\subsection{Incorporating the Distribution Function}

As $q^2$ is lowered below $(m_B-\mxcut)^2$ the effects of the distribution
function become progressively more important, and their size becomes a detailed
question depending on the difference between the left- and right-hand sides of
Eq.~(\ref{qsqblowup}).  The region where the distribution function becomes
significant is  correlated with the region where the Sudakov logs from the
singularity  (\ref{qsqblowup}) get large.  In the simple model discussed in this
section, the impact of the distribution  function on the partially integrated
rate is indeed roughly constant along the thin  dashed lines in
Fig.~\ref{singularityplot}, independent of the value of $\mxcut$.

The $b$ quark light-cone distribution function  can be measured from the shape
of the photon spectrum in $B\to X_s\gamma$, but  in the near future such a
measurement will have sizable experimental  uncertainties. There are also
unknown $O(1/m_b)$ corrections in relating this function  to the one relevant
for semileptonic $B$ decay (see \cite{blm01} for a  discussion of these terms in
the twist expansion).   In this paper we restrict  ourselves to cuts for which
the effect of the distribution function is small, so that its measurement error
and the unknown $O(1/m_b)$ corrections have a small effect in the determination
of $|V_{ub}|$.

We still need to estimate the effect of the distribution function to  determine
how low $\qcut^2$ may be decreased.  Since we restrict ourselves to regions
where the effect of the structure function is small, it is sufficient to take
them into account at tree level. To leading twist, this is obtained by smearing
the $b$ quark decay rate with the distribution function $f(k_+)$, which amounts
to the replacement in Eq.~(\ref{q2spec}),
\begin{equation}
\delta(\shat) \to \cases{
  \displaystyle \int \!{\rm d} k_+\,\delta \bigg(\shat + \frac{1-\qsh}{m_b}\,
k_+ \bigg) f(k_+), &~for $\hat q^2<(1-\hat\mxcut)^2$ \vspace*{6pt}\cr
  \displaystyle \delta(\shat), & otherwise.\cr}
\end{equation}
(We do not include the distribution function in the region $\hat
q^2>(1-\hat\mxcut)^2$, since in this region its effects are contained in the
$O(\lqcd^3/m_b^3)$ terms, which we have already considered.) This corresponds
to multiplying the leading order result Eq.~(\ref{singlecut}) in the region
$\hat q^2<(1-\hat\mxcut)^2$ by a factor
\begin{eqnarray}\label{Afunct}
A(\qsh,\mxcut) = \int_{-s_0\frac{m_b}{1-\qsh}}^{\Lambda}
  \!{\rm d} k_+\, f(k_+) \,,
\end{eqnarray}
where $s_0$ is defined in Eq.~(\ref{s0limit}) and $\Lambda\equiv
m_B-m_b.$\footnote {Since there are order $\lqcd/m_b$ corrections to the
distribution function, we do not need to distinguish between $\Lambda$  and the
HQET parameter $\bar\Lambda$.}
The best way to determine $f(k_+)$ is from the $B\to X_s\gamma$ photon
spectrum, which gives at tree level
\begin{eqnarray}
A(\qsh,\mxcut) = \frac1{2\Gamma^\gamma K}\int_
{-s_0\frac{m_b}{1-\qsh}}^{\Lambda}
  {\rm d}k_+ \left.
  {{\rm d}\Gamma^\gamma \over {\rm d}E_\gamma}\right|
  _{E_\gamma = \frac{m_b+k_+}2 }
\end{eqnarray}
where $K\sim 1.33$ takes into account contributions from operators other than
$O_7$ to the photon spectrum \cite{leptonendpoint2}, and $\Gamma^\gamma$ is the
contribution of the tree level matrix element of $O_7$ to the $B\to X_s\gamma$
decay rate. Thus the experimental data on the $B\to X_s\gamma$ photon energy
spectrum will make the estimate of this source of error small and largely model
independent. (Note that the result is modified by large Sudakov logs, which in
principle should be resummed, but in the region we are interested these effects
are subleading and may be neglected.) Since the dependence of our results on
$f(k_+)$ is weak, even a crude measurement will facilitate a model independent
determination of $|V_{ub}|$ from the combined $q^2$ and $m_X$ cuts with small
errors.

In the absence of precise data, we will use the simple model presented in
\cite{defazio} to estimate the effects of the structure function,
\begin{eqnarray}\label{structmodel}
f(k_+) = {a^a\over\Gamma(a)}\, (1-x)^{a-1}\, e^{-a(1-x)}\,, \quad
   x = \frac{k_+}{\Lambda}\,, \quad
   a = -{3\Lambda^2\over \lambda_1} \,.
\end{eqnarray}
This model is chosen such that its first few moments satisfy the known
constraints: the zeroth moment (with respect to $x$) is unity, the first moment
vanishes, and the second moment is $\lambda_1/3m_b^2$.

\begin{figure}[t]
\centerline{\includegraphics[width=4in]{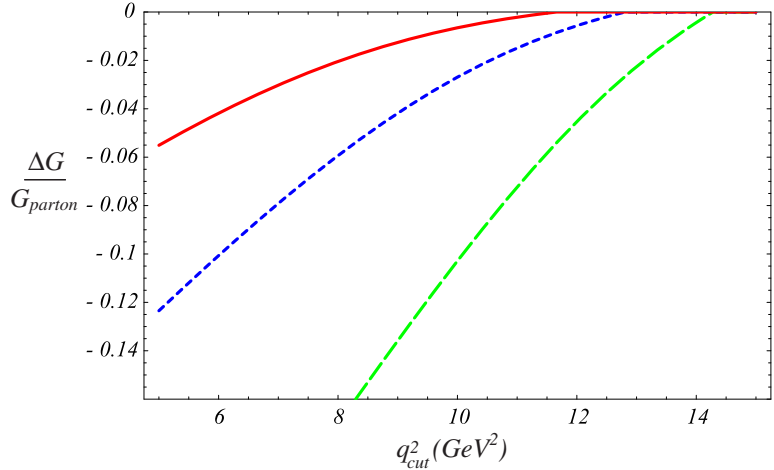}}
\caption{The effect of the model structure function (\ref{structmodel})
on $\gcut$ as a function of
$\qcut^2$ for $\mxcut = 1.86\,\gev$ (solid line), $1.7\,\gev$ (short
dashed line) and $1.5\,\gev$ (long dashed line).}
\label{Sfig}
\end{figure}

In Fig.~\ref{Sfig} we plot in this model the effect of the structure function
on $\gcut$ as a function of $\qcut^2$, for three different values of  $\mxcut$.
The curves correspond to the parameters $\Lambda=0.57\,\gev$ and
$\lambda_1=-0.2\,\gev^2$.

\section{Combined Results}\label{results}

\begin{figure}
\centerline{\includegraphics[width=4in]{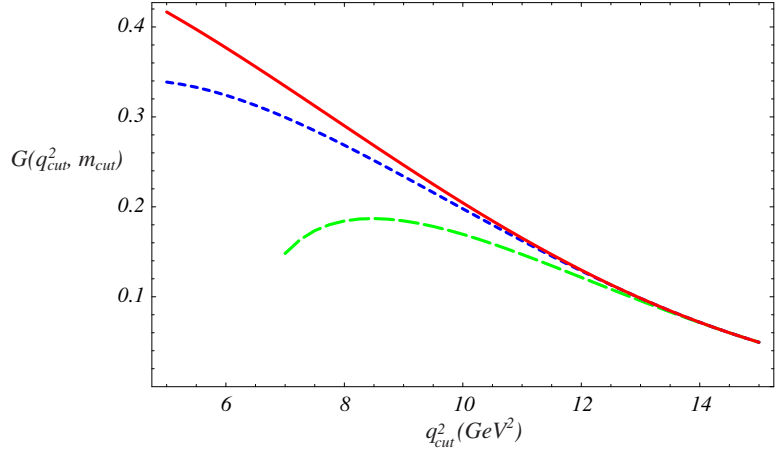}}
\caption{$\gcut$, which determines the partially integrated rate according to
Eq.~(\ref{defineg}), as a function of the dilepton invariant mass cut
$\qcut^2$, for hadronic invariant mass cut $\mxcut=1.86\,\gev$ (solid  line),
$1.7\,\gev$ (short dashed line) and $1.5\,\gev$ (long dashed line). }
\label{final}
\end{figure}

Having considered each uncertainty separately, we now combine them and  give
the final result for various values of cuts $(\qcut^2, \mxcut)$. In
Fig.~\ref{final} we plot $\gcut$ as a function of $\qcut^2$ for  $\mxcut =
1.5\,\gev$, $1.7\,\gev$ and $1.86\,\gev$.  In this figure we choose the  values
$\mbups = 4.7\, \gev$, $\lambda_1 = -0.2\, \gev^2$ and $\alpha_s(m_b) = 0.22$.
The combined cut on $q^2$ and $m_X$ allows a  determination of
$|V_{ub}|$ from about twice the fraction of events than in the case of the cut
on $q^2$ alone. The turnaround of the curve for $\mxcut = 1.5\,\gev$ signals
the breakdown of the perturbation expansion due to the singularity at $\hat
s=0$, and is not physical.

\begin{table}[ht]
\begin{tabular}{c||c|c|ccc|c}
Cuts on $(q^2,\,m_X^2)$  &  ~$\gcut$~
   &  ~$\Delta_{\rm struct}G$~ &  ~$\Delta_{\rm pert}G$  &  
   $\matrix{\Delta_{m_b}G \cr {\footnotesize \pm80/30\,\mev}}$  &
   $\Delta_{1/m^3} G$~   &  $\Delta G$  \\ \hline\hline
\multicolumn{1}{c}{Combined cuts} & \multicolumn{6}{c}{} \\ \hline
$6\,\gev^2, 1.86\,\gev$	&  0.38  &  $ -4\%$  &4\%  &  13\%/5\%  &  6\%  &
15\%/9\%  \\
$8\,\gev^2, 1.7\,\gev$ 	&  0.27 &  $-6\%$&  6\%  &  15\%/6\%  &  8\%  &
18\%/12\% \\ 
$ 11\,\gev^2, 1.5\,\gev$  & 0.15 &  $-7\%$ &13\% & 18\%/7\% & 16\% &
27\%/22\% \\ \hline\hline
\multicolumn{1}{c}{Pure $q^2$ cuts} & \multicolumn{6}{c}{} \\ \hline
~$(m_B-m_D)^2, m_D$~ & 0.14 & --\,--&15\% &19\%/7\% & 18\% & 
~30\%/24\%~ \\
$(m_B-m_{D^*})^2, m_{D^*}$& 0.17 & --\,--&13\% &17\%/7\% & 14\% &26\%/20\%
\end{tabular}\vspace*{4pt}
\caption{$\gcut$, as defined in Eq.~(\ref{defineg}), for several different choices 
of $(\qcut^2,\mxcut)$, along with the uncertainties.  The
fraction of $B\to X_u\ell\bar\nu$ events included by the cuts is $1.21\,\gcut$.
The two last lines corresponding to pure $q^2$ cuts
are included for comparison.  $\Delta_{\rm
struct}G$ gives the fractional  effect of the structure function $f(k_+)$ in
the simple model (\ref{structmodel}); we do not include an uncertainty on
this in our error estimate. The overall uncertainty $\Delta G$ is obtained
by combining the other uncertainties in quadrature.   The two values correspond
to $\Delta \mbups=\pm 80\,\mev$ and $\pm 30\,\mev$. The uncertainty in
$|V_{ub}|$ is half of $\Delta G$. }
\label{finaltable}
\end{table}

In Table~\ref{finaltable} we use three representative sets of cuts 
in $q^2$ and $m_X$ to estimate
the overall theoretical uncertainty with which $|V_{ub}|$ can be determined. As
throughout this paper, we choose for the cut on the hadronic invariant mass the
three values $\mxcut = (1.5\,\gev,\, 1.7\,\gev,\, 1.86\,\gev)$.   We choose
values of $\qcut^2$ which keep the effects of the distribution function $f(k_+)$
small (in the simple model discussed in the previous section).  Because we
anticipate the distribution function will be extracted from the $B\to X_s\gamma$
spectrum to the accuracy required, we do not include an uncertainty on $f(k_+)$
in our overall theoretical uncertainty.

For comparison, we include in Table~\ref{finaltable} the results for a pure
$q^2$ cut (corresponding to $\mxcut=m_B-\sqrt{\qcut^2}$), for $q^2=(m_B-
m_D)^2=11.6\,\gev^2$ and $q^2=(m_B-m_{D^*})^2=10.7\,\gev^2$.  We include the
second point because $B\to D\ell\bar\nu_\ell$ is suppressed near zero recoil,
and so may be reliably subtracted from the background \cite{BLL2}.  These
results are consistent with \cite{neubbech}, with comparable errors from
perturbation theory and $m_b$ variation.  

A source of uncertainty not explicitly considered in this paper arises from 
possible quark-hadron duality violation.  The size of this is difficult to
estimate theoretically, but based on the agreement the values of $|V_{cb}|$
extracted from inclusive and exclusive $B$ decays, we expect it to be smaller
than the uncertainties we have considered.  Cuts on the phase space may amplify
duality violation, but since this technique may be sensitive to almost half of
the events, we expect these effects to remain small.  In any event, this can be
tested experimentally by comparing the extraction of $|V_{ub}|$ with different
values of the cuts. 

Ultimately, experimental considerations will determine the optimal values of
$(\qcut^2, \mxcut)$. An actual analysis will probably be sensitive to the
region $q^2 > \qcut^2$ and $m_X < \mxcut$  with non-uniform weight.  The
theoretical errors in such a case will be  comparable to our results, as long
as the weight function does not vary too  rapidly.  The formulae presented in
the Appendix are sufficient to determine the perturbative relationship of
$|V_{ub}|$ and such a measurement. In addition, as explained in \cite{BLL2},
due to heavy quark symmetry, the  $B\to X_c\ell\bar\nu$ background near $m_X =
m_D$ may be easier to understand  as a function of $q^2$ and $m_X$ than as a
function of $m_X$ only. For  example, the $D^{**}$ and higher mass states
cannot contribute for $q^2 > 8.5\,  \gev^2$, and so the main background is
$B\to D^* \ell\bar\nu$ near zero recoil, which will be precisely measured to
determine $|V_{cb}|$.

\section{Conclusions}

In this paper we proposed a precision determination of the magnitude of the CKM
matrix element $V_{ub}$ from charmless inclusive semileptonic $B$ decays using
combined cuts on the dilepton invariant mass, $q^2$, and the hadronic invariant
mass, $m_X$.  This leads to the following general strategy for determining
$|V_{ub}|$:
\begin{itemize}
\item make the cut on $m_X$ as large as possible, keeping the background from 
$B$ to charm under control 
\item for a given cut on $m_X$, reduce the $q^2$ cut as low as possible,
keeping the contribution from the $b$ quark structure function, as well as the
perturbative uncertainties, small (see Figs.\ \ref{perturbative} and
\ref{Sfig}).  
\end{itemize}
We have calculated $\gcut$, the partially integrated rate in the presence of
cuts on $q^2$ and $m_X$ (normalized as in Eq. (\ref{defineg})).  Our results
are summarized for three representative values of the cuts in Table
\ref{finaltable}. The total uncertainty $\Delta G$ is twice the uncertainty in
$|V_{ub}|$. The uncertainty from weak annihilation (Fig.\ \ref{mcubed}) may be 
reduced by comparing results in $B^\pm$ and $B^0$ decay, or by comparing the
semileptonic widths of the $D^0$ and $D_s$ \cite{Voloshin}, while the remaining
uncertainties could be reduced by an improved determination of the $b$ quark
mass and a complete two loop calculation of the doubly differential rate
$\d\Gamma/\d q^2\,\d m_X$.  

This method is sensitive to up to $\sim 45\%$ of the $B\to X_u\ell\nu$ decays,
about twice the fraction of events than in the case of the cut on $q^2$ alone.
We found that a determination of $|V_{ub}|$ with a theoretical error at the
5--10\% level is possible.  The combined $(\qcut^2, \mxcut)$ cut also allows
this precision to be obtained with cuts which are away from the threshold for
$B\to X_c\ell\bar\nu_\ell$, an important criterion for realistic detector
resolution. Such a measurement of $|V_{ub}|$ would  largely reduce the standard
model range of $\sin2\beta$, and thus allow more sensitive searches for new
physics.

\acknowledgements

We thank Adam Falk, Lawrence Gibbons and Ian Shipsey for helpful discussions. 
This work was supported in part by the Natural Sciences and Engineering
Research Council of Canada. Z.L.~was supported in part by the Director, Office
of Science, Office of High Energy and Nuclear Physics, Division of High Energy
Physics, of the U.S.\ Department of Energy under Contract DE-AC03-76SF00098.
C.B. was supported by the US Department of Energy under contract
DE-FG03-97ER40546.
C.B. thanks the theory group at LBL for its hospitality while some of this work
was completed.

\appendix

\section*{The functions $\widetilde X(\qsh,\mxcut)$ and
$\widetilde Y(\qsh,\mxcut)$}

The functions $\xcut(\qsh, \mxcut)$ and $\ycut(\qsh, \mxcut)$ in
Eq.~(\ref{singlecut}) can be determined from $X(\qsh, \shat)$ and  $Y(\qsh,
\shat)$ defined in Eq.~(\ref{q2spec}) via
\begin{equation}
\xcut(\qsh,\mxcut) = \int_0^{s_0} {\rm d}\shat\, X(\qsh, \shat) \,,
   \quad
\ycut(\qsh,\mxcut) = \int_0^{s_0} {\rm d}\shat\, Y(\qsh, \shat) \,,
\end{equation}
where $s_0$ is given in (\ref{s0limit}).

When $\mxcut > m_B - m_b\,\sqrt{\qsh}$, the $\mxcut$ limit does not  restrict
the ${\rm d}\shat$ integration, and the result is just the value of the  single
differential $\qsh$ spectrum.  The order $\alpha_s$ correction to ${\rm
d}\Gamma/{\rm d}\qsh$ was computed in Ref.~\cite{JK},
\begin{eqnarray}
\tilde X_0(\qsh) = - \frac23\, && \bigg\{
   2(1-\qsh)^2 (1+2\qsh) \Big[ \pi^2 + 2 L_2(\qsh) - 2 L_2(1-\qsh) \Big]
   + 4 \qsh (1-\qsh-2\qh^4) \ln\qsh \nonumber\\
&&{} + 2 (1-\qsh)^2 (5+4\qsh) \ln(1-\qsh) - (1-\qsh) (5+9\qsh-6\qh^4)
   \bigg\} \,,
\end{eqnarray}
where $L_2(z) = -\int_0^z {\rm d}t \ln(1-t)/t$ is the dilogarithm. The order
$\alpha_s^2\beta_0$ correction to ${\rm d}\Gamma/{\rm d}\qsh$  was computed in
Ref.~\cite{LSW} numerically.  We find that the following  simple function
\begin{equation}
\tilde Y_0(\qsh) \simeq 0.472\, (1-\qsh) - 32.5\, (1-\qsh)^2
   + 42.3\, (1-\qsh)^3 - 16.0\, (1-\qsh)^4 \,,
\end{equation}
gives a very good approximation.  It deviates from the exact result by  less
than $0.01$ for any value of $\qsh$ (while $\int_0^1 \tilde Y_0(\qsh)\, {\rm d}
\qsh \simeq -3.22$).

In the second case in Eq.~(\ref{s0limit}), $\mxcut^2 < (m_B-m_b\,\qsh)\,
(m_B-m_b)$,  $\mxcut$ is too small, and the perturbative calculation is  not
reliable.  As we have discussed, we avoid this region in
this paper.

The situation in which neither of the first two cases in Eq.~(\ref{s0limit})
applies is the  most interesting for us. We obtain
\begin{eqnarray}
\xcut(\qsh, \mxcut) &=& \tilde X_0(\qsh) - \frac43\, (1-\qsh)^2 (1+2\qsh)
  \Bigg\{ \frac{\pi^2}3 - \frac72 \ln(4s_0)  + 2 (\ln 2)^2 - (\ln s_0)^2 \\*
&& +2 \ln T \ln\frac{4s_0^2}T
  + \left(3 + 2\ln\frac{T - R + s_0}{4T^2} \right) \ln(T-R+s_0)  \nonumber\\*
&& - 2 L_2(T) + 4 L_2 \bigg( \frac{T + R + s_0}2 \bigg)
  - 4 L_2 \bigg( \frac{T + R + s_0}{2T}\bigg) \Bigg\} \nonumber\\*
&& - \frac43\, \Bigg\{ R (5+7\qsh-8\qh^4+s_0)
  + s_0\, (1+2\qsh)\, (4T + s_0) \ln\frac{T-R+s_0}{2\sqrt{s_0}} \nonumber\\*
&& - 4 \qsh (1+\qsh) (1-2\qsh) \ln\frac{T-R-s_0}{\sqrt{\qsh}} 
  + 4(1+\qsh-4\qh^4) \ln(T-R+s_0) \Bigg\} \,, \nonumber
\end{eqnarray}
where $R = \sqrt{\qh^4+(1-s_0)^2-2\qsh(1+s_0)}\,$, $T=1-\qsh$,
and $s_0$ is given in Eq.~(\ref{s0limit}).
For the coefficient of the order $\alpha_s^2\beta_0$ correction we find
\begin{eqnarray}
\ycut(\qsh, \mxcut) = \tilde Y_0(\qsh) -
   \int_{s_0}^{\left(1-\sqrt{\qsh}\right)^2} {\rm d}s\, \frac12\,
   Z_2\bigg(s,\frac{1+s-\qsh}2 \bigg) \,,
\end{eqnarray}
where
\begin{eqnarray}
Z_2(s,e) &=& \bigg( \frac5{12} - \frac14 \ln s \bigg) Z_1(s,e)
   - \frac2{3s}\, \Bigg\{
   \sqrt{e^2-s}\, \Big[ 5 e (3 - 4 e) - 4 s + 26 e s - 8 s^2 \Big]  \\*
&&{} + s\, \bigg[9 - 9 e - 8 e^2 + \frac{8e^3}s + s + 6 e s - 2 s^2
     - \frac{3(1-e)}{1-2e+s} \bigg] \ln\frac{(e + \sqrt{e^2 - s})^2}s
\nonumber\\*
&&{} + \frac{(2 e - s)\sqrt{e^2-s}}{1 - 2 e + s}\,
    \Big[12 + 40 e^2 + 5 s (5 + 2 s) - 4 e (11 + 10 s) \Big]
    \ln\frac{(2 e - s)^2}s  \nonumber\\*
&&{} + (3 - 4 e + 2 s) (8 e^2 - 4 e s + s^2)
    \Bigg[ L_2\Bigg( \frac{\sqrt{e^2-s} + e - s}{\sqrt{e^2 - s} - e}
\Bigg)
    - L_2\Bigg( \frac{\sqrt{e^2-s} - e + s}{\sqrt{e^2 - s} + e} \Bigg)
\Bigg]
   \Bigg\} \nonumber
\end{eqnarray}
and
\begin{eqnarray}
Z_1(s,e) &=& \frac{16}{3s}\, \sqrt{e^2 - s}\,
   [28 e^2  + 2 s (5 + 4 s) - 3 e (7 + 10 s)] \nonumber\\*
&& - \frac8{3s}\, (3 - 4 e + 2 s) (8 e^2 - 4 e s + s^2)
   \ln\frac{e - \sqrt{e^2 - s}}{e + \sqrt{e^2 - s}} \,.
\end{eqnarray}

%%%%%%%%%%%%%%%%%%%%%%%%%%%%%%%%%%%%%%%%%%
%Bibliography


\begin{references}

\bibitem{OPE}
J.~Chay, H.~Georgi and B.~Grinstein, Phys.\ Lett.\ B247 (1990) 399; \\
%%CITATION = PHLTA,B247,399;%%
%
M.A.~Shifman and M.B.~Voloshin, Sov.\ J.\ Nucl.\ Phys.\ 41 (1985) 120; \\
%%CITATION = YAFIA,41,187;%%
%
I.I.~Bigi {\it et al.}, Phys.\ Lett.\ B293 (1992) 430
[Erratum-ibid.\ B 297, 430 (1992)]; \\
%%CITATION = HEP-PH 9207214;%%
%
I.I.~Bigi {\it et al.}, Phys.\ Rev.\ Lett.\ 71 (1993) 496; \\
%%CITATION = HEP-PH 9304225;%%
%
A.V.~Manohar and M.B.~Wise, Phys.\ Rev.\ D49 (1994) 1310.
%%CITATION = HEP-PH 9308246;%%

\bibitem{Elexp}
F. Bartelt {\it et al.}, CLEO Collaboration, Phys. Rev. Lett. 71 (1993) 4111;\\
%%CITATION = PRLTA,71,4111;%%
H. Albrecht {\it et al.}, Argus Collaboration, Phys. Lett. B255 (1991) 297.
%%CITATION = PHLTA,B255,297;%%

\bibitem{mXcut}
A.F.~Falk, Z.~Ligeti and M.B.~Wise, Phys.\ Lett.\ B406 (1997) 225; \\
%%CITATION = HEP-PH 9705235;%%
%
I.~Bigi, R.~D.~Dikeman and N.~Uraltsev, Eur.\ Phys.\ J.\ C4 (1998) 453 .
%%CITATION = HEP-PH 9706520;%%

\bibitem{mXold}
V. Barger {\it et al.}, Phys. Lett. B251 (1990) 629;\\
%%CITATION = PHLTA,B251,629;%%
J. Dai, Phys. Lett. B333 (1994) 212.
%%CITATION = HEP-PH 9405270;%%

\bibitem{mXexp}
R. Barate {\it et al.}, ALEPH Collaboration, CERN EP/98-067;
DELPHI Collaboration, contributed paper to the ICHEP98 Conference
(Vancouver), paper 241;\\
M. Acciarri {\it et al.}, L3 Collaboration, Phys. Lett. B436 (1998) 174.
%%CITATION = PHLTA,B436,174;%%

\bibitem{BLL}
C.W.~Bauer, Z.~Ligeti and M.~Luke, Phys.\ Lett.\ B479 (2000) 395.
%%CITATION = HEP-PH 0002161;%%

\bibitem{BLL2}
C.W.~Bauer, Z.~Ligeti and M.~Luke, hep-ph/0007054.
%%CITATION = HEP-PH 0007054;%%


\bibitem{Voloshin}
M.B.~Voloshin, hep-ph/0106040.
%%CITATION = HEP-PH 0106040;%%

\bibitem{structure}
M.~Neubert, Phys.\ Rev.\ D49 (1994) 4623; \\
%%CITATION = HEP-PH 9312311;%%
%
I.I.~Bigi {\it et al.}, Int.\ J.\ Mod.\ Phys.\ A9 (1994) 2467.
%%CITATION = HEP-PH 9312359;%%

\bibitem{extractshape}
A.K.~Leibovich, I.~Low and I.Z.~Rothstein, Phys.\ Rev.\ D61 (2000) 053006;\\
%%CITATION = HEP-PH 9909404;%%
%
%A.~K.~Leibovich, I.~Low and I.~Z.~Rothstein,
Phys.\ Lett.\ B486 (2000) 86.
%%CITATION = HEP-PH 0005124;%%


\bibitem{neubertq2}
M.~Neubert, JHEP 0007 (2000) 022.
%%CITATION = HEP-PH 0006068;%%

\bibitem{defazio}
F.~De Fazio and M.~Neubert, JHEP 9906 (1999) 017.
%%CITATION = HEP-PH 9905351;%%

\bibitem{LSW}
M. Luke, M. Savage, and M.B. Wise, Phys. Lett. B343 (1995) 329.
%%CITATION = HEP-PH 9409287;%%

\bibitem{upsexp}
A.H. Hoang, Z. Ligeti, and A.V. Manohar, Phys. Rev. Lett. 82 (1999) 277;
%%CITATION = HEP-PH 9809423;%%
Phys. Rev. D59 (1999) 074017.
%%CITATION = HEP-PH 9811239;%%

\bibitem{neubbech}
M.~Neubert and T.~Becher,
%``Improved determination of $|$V(ub)$|$ from inclusive semileptonic B-meson
% decays,''
hep-ph/0105217.
%%CITATION = HEP-PH 0105217;%%

\bibitem{hqetlogs}
M.~A.~Shifman and M.~B.~Voloshin,
Sov.\ J.\ Nucl.\ Phys.\  {45}, 292 (1987);\\
%%CITATION = SJNCA,45,292.1987\ YAFIA,45,463;%%
H.~D.~Politzer and M.~B.~Wise,
Phys.\ Lett.\ B{206}, 681 (1988);\\
%%CITATION = PHLTA,B206,681;%%
A.~F.~Falk and B.~Grinstein,
Phys.\ Lett.\ B{247}, 406 (1990);\\
%%CITATION = PHLTA,B247,406;%%
A.~Czarnecki,
%``Two-loop QCD corrections to $b\to c$ transitions at zero recoil,''
Phys.\ Rev.\ Lett.\  {76}, 4124 (1996).
%%CITATION = HEP-PH 9603261;%%

\bibitem{sumrules1}
M.~B.~Voloshin, Int.\ J.\ Mod.\ Phys.\ A10 (1995) 2865;\\
%%CITATION = HEP-PH 9502224;%%
A.~A.~Penin and A.~A.~Pivovarov, Phys.\ Lett.\ B435 (1998) 413;\\
%%CITATION = HEP-PH 9803363;%%
K.~Melnikov and A.~Yelkhovsky, Phys.\ Rev.\ D59 (1999) 114009;\\
%%CITATION = HEP-PH 9805270;%%
A.~H.~Hoang, Phys.\ Rev.\ D61 (2000) 034005.
%%CITATION = HEP-PH 9905550;%%

\bibitem{benekesigner}
M.~Beneke and A.~Signer, Phys.\ Lett.\ B471 (1999) 233.
%%CITATION = HEP-PH 9906475;%%

\bibitem{hoang00}
A.~H.~Hoang, hep-ph/0008102.
%%CITATION = HEP-PH 0008102;%%

\bibitem{decaydists}
M.B.~Voloshin, Phys.\ Rev.\ D51 (1995) 4934;\\
%%CITATION = HEP-PH 9411296;%%
A.~Kapustin and Z.~Ligeti, Phys.\ Lett.\ B355 (1995) 318;\\
%%CITATION = HEP-PH 9506201;%%
M. Gremm, A. Kapustin, Z. Ligeti, and M.B. Wise, 
Phys. Rev. Lett. 77 (1996) 20;\\
%%CITATION = HEP-PH 9603314;%%
A.~F.~Falk, M.~Luke and M.~J.~Savage, Phys.\ Rev.\ D53 (1996) 6316;
%%CITATION = HEP-PH 9511454;%%
Phys. Rev. D53 (1996) 2491;\\
%%CITATION = HEP-PH 9507284;%%
Z.~Ligeti, M.~Luke, A.V.~Manohar and M.B.~Wise, Phys.\ Rev.\ D60 (1999)
034019.
%%CITATION = HEP-PH 9903305;%%

\bibitem{m3}
C.~W.~Bauer and C.~N.~Burrell, Phys.\ Rev.\ D62 (2000) 114028; \\
%%CITATION = HEP-PH 9911404;%%
M.~Gremm and A.~Kapustin, Phys.\ Rev.\ D55 (1997) 6924.
%%CITATION = HEP-PH 9603448;%%

\bibitem{leptonendpoint1}
A.~K.~Leibovich and I.~Z.~Rothstein,
%``The resummed rate for B $\to$ X/s gamma,''
Phys.\ Rev.\ D 61, 074006 (2000);\\
%%CITATION = HEP-PH 9907391;%%
A.~K.~Leibovich, I.~Low and I.~Z.~Rothstein,
%``A comment on the extractions of V(ub) from radiative decays,''
hep-ph/0105066.
%%CITATION = HEP-PH 0105066;%%

\bibitem{leptonendpoint2}
M.~Neubert,
%``Note on the extraction of $|$V(ub)$|$ using radiative B decays,''
hep-ph/0104280;
%%CITATION = HEP-PH 0104280;%%

\bibitem{blm01}
C.~W.~Bauer, M.~Luke and T.~Mannel, hep-ph/0102089.
%%CITATION = HEP-PH 0102089;%%

\bibitem{JK}
M.~Jezabek and J.~H.~Kuhn,
Nucl.\ Phys.\ B314 (1989) 1.
%%CITATION = NUPHA,B314,1;%%


\end{references}
\end{document}